\documentclass[%
 reprint,
superscriptaddress,
 amsmath,amssymb,
 aps,
 prb,
floatfix,
]{revtex4-2}

\usepackage{graphicx}
\usepackage{amsmath}
\usepackage{dcolumn}
\usepackage{bm}
\usepackage{physics}
\usepackage{float}
\usepackage{tabularx}
\usepackage{multirow}
\usepackage{gensymb}
\usepackage{textgreek}
\usepackage{booktabs}
\usepackage{array}
\usepackage[caption=false]{subfig}
\usepackage{color,soul}

\begin{document}


\title{Effect of Biaxial Strain on Cation Octahedral Rotations and \textcolor{black}{Magnetic} Structure of the Antiperovskite Mn\textsubscript{3}GaN}

\author{Roman Malyshev}
\affiliation{Department of Electronic Systems, Norwegian University of Science and Technology, NO-7491 Trondheim, Norway}
\author{Ingeborg-Helene Svenum}
\affiliation{SINTEF Industry, NO-7465 Trondheim, Norway}
\affiliation{Department of Chemical Engineering, Norwegian University of Science and Technology, NO-7491 Trondheim, Norway}
\author{Sverre M. Selbach}
\affiliation{Department of Materials Science and Engineering, Norwegian University of Science and Technology, NO-7491 Trondheim, Norway}
\author{Thomas Tybell}
\affiliation{Department of Electronic Systems, Norwegian University of Science and Technology, NO-7491 Trondheim, Norway}

\date{\today}

\begin{abstract}
Density functional theory is used to study the effect of compressive and tensile biaxial strain on Mn\textsubscript{3}GaN. Mn\textsubscript{3}GaN is a non-collinear antiferromagnetic antiperovskite with a similar structure to that of an ideal cubic oxide perovskite, but with cations at the octahedral sites while the anion, nitrogen, is found at the B site. The present study explores the response of Mn\textsubscript{3}GaN to (001) strain, considering biaxial strain levels ranging from -5\% to 5\%. It is found that the electron structure is insensitive to tensile strain. The study supports previous results in that a spin-canted antiferromagnetic order emerges due to tensile strain, inducing net magnetization. Compressive strain collapses the non-collinear antiferromagnetic spin structure and induces a ferrimagnetic order at -2\% strain. Notably, in contrast with oxide perovskites, Mn\textsubscript{3}GaN does not respond to strain by octahedral tilt, but rather by intraband redistributions of charge between Mn $d$ states. Despite the similar structure to oxide perovskites, the bonds between the B site anion and octahedral site cations in Mn\textsubscript{3}GaN bonds are less rigid, such that strain is instead accommodated by a change in bond length rather than a change in bond angles.

\end{abstract}

\maketitle

\section{\label{sec:introduction}Introduction}
\begin{figure}[ht]
   \centering
   \includegraphics[width=0.9\linewidth]{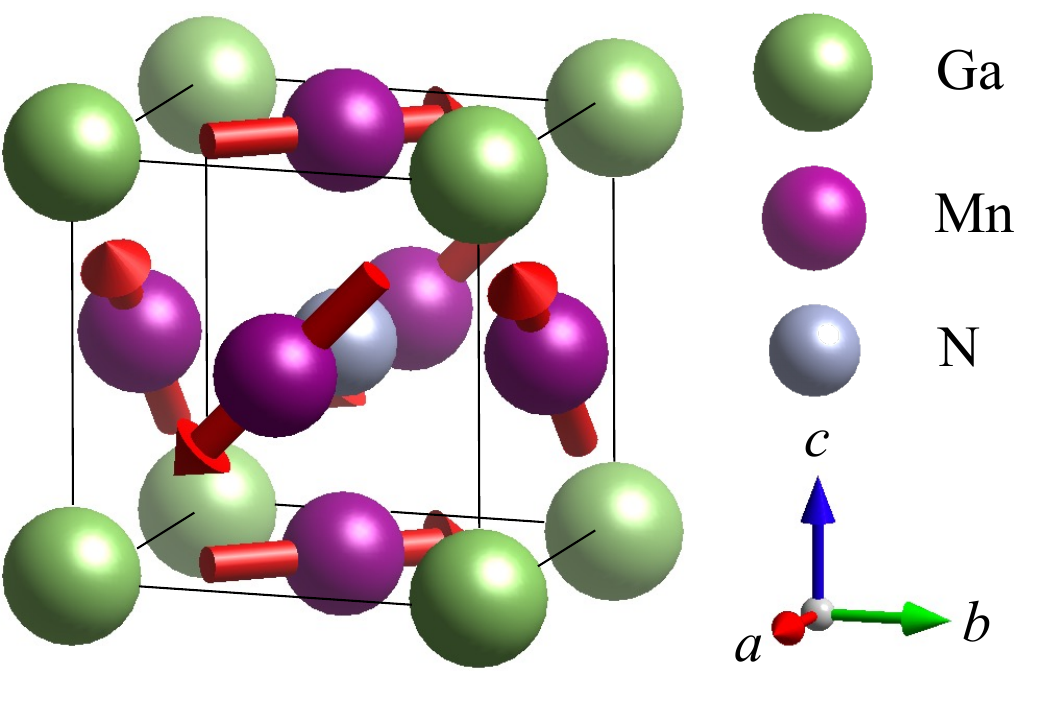}
    \caption{MGN unitcell with spin vectors illustrating its $\Gamma^\text{5g}$ spin structure. The scalar values of the spin vectors are not to scale.}
    \label{fig:MGN_unitcell}
\end{figure}

Antiperovskites are intermetallic compounds that have been shown to exhibit functional properties, including giant magnetoresistance \cite{Mn3GaC_GMR_Kamishima_PRB, Mn3GaC_Ni_GMR_APL_2009}, superconductivity \cite{MgCNi3_superconduct_nature_2001, CuNNi3_supercond_2013, CdCNi_supercond_2007, LaPd3P_supercond_2021},
thermoelectricity \cite{ANE_Mn3SnN, thermoel_antiperov} and piezomagnetism \cite{piezomagn_Mn, piezomagn_Mn3NiN}. In particular, manganese nitrides exhibit non-collinear antiferromagnetic (AFM) spin structures \cite{MGN_neg_therm_expansion, MGN_bulk_TN, manganese_nitride_spin_polarization}. Mn\textsubscript{3}GaN (MGN) has a Kagome-like $\Gamma^\text{5g}$ noncollinear antiferromagnetic spin structure along the (111) plane \cite{MGN_lukashev, MGN_magnetism_caloric_effects}, illustrated in Figure \ref{fig:MGN_unitcell}, having a zero net magnetic moment per unit cell in bulk material. MGN has been studied for\\ its magnetic bulk properties, but also in magnetoelectric heterostructures, demonstrating spin-orbit torque switching \cite{Electrical_current_switchingMn3GaN_hajiri_ishino, nan_quintela_nature_comm}.

A unit cell of the cubic antiperovskite MGN, space group \textit{Pm$\bar{3}$m}, is structured similarly to a cubic perovskite such as SrTiO3 \cite{sto_1964}, but with a cation octahedron Mn\textsubscript{6}N surrounding a center anion, rather than an anion octahedron and a center cation. The lattice constant in bulk material measures 3.886 \AA\, at 0\degree C \cite{mgn_consts}. At $T_N = 345-355$ K it undergoes a first-order antiferromagnetic-paramagnetic transition with negative thermal compressibility. It compresses by approximately 1.1 \% cell volume \cite{MGN_bulk_TN, MGN_neg_therm_expansion, nan_quintela_nature_comm}. Spin-polarized ferri- and ferromagnetic orders have been observed in MGN below room temperature at 10-50 K \cite{mn3gan_spin_glass, mn3gan_ptype} and ferromagnetism has been predicted for MGN in its ground state \cite{mn3ga_volume_strain, MGN_magnetism_caloric_effects}. Changes in stoichiometry, in particular nitrogen deficiency, Mn\textsubscript{3}GaN\textsubscript{1-x}, can produce tetragonal distortions resulting in ferromagnetism at room temperature with Curie temperatures at 660-740 K \cite{mn3gan1-x_ishinoa_so, Electrical_current_switchingMn3GaN_hajiri_ishino, ferromagnetic_mgn_lee_sukegawa_2015}. \\

The observation of spin-polarized phases in stoichiometric MGN at low temperature raises a question, whether there is a structure-property relation that can be exploited to invoke the different phases. In perovskite\\ oxides, a relatively flat energy landscape allows the use of strain-engineering to stabilize functional phases that would normally be energetically unfavorable \cite{adv_mat_spaldin_rondinelli_2011}, such as transitioning between antiferromagnetism and ferromagnetism \cite{martinsen_amo3_2016_mrs_comm, SrMnO3_multiferroic_strain, LSMOstrainDFT2000, LSMOstrain1999}, inducing a ferroelectric state or enhancing a polar phase \cite{sto_strained_ferroel, batio3_ferroics_strain, batio3_enhance_ferroelectric_strain, smo_strain_nature_2015}. The similarity of the antiperovskite structure to that of oxide perovskites opens for adopting similar approaches to control the electronic ground state, including strain-engineering.

A density functional theory study is presented here on how the electronic and magnetic structures of cubic MGN are affected by epitaxial (001) strain. It is shown that biaxial strain does not induce octahedral rotations typical for oxide perovskites, rendering the material tetragonal with no octahedral tilt ($P4/mmm$). Strain does, however, alter the electronic structure of the material and the magnetic order. It is shown that compressive strain can induce a magnetic phase transition, transforming the antiferromagnetic non-collinear order into a spin-polarized ferrimagnetic structure. Tensile strain, however, preserves a Kagome-like spin lattice, but introduces spin canting, yielding a net magnetization.

\section{\label{sec:methodology}Computational Method}
The MGN structures were simulated using density functional theory as implemented in the \textit{Vienna Ab Initio Simulation Package} (\textit{VASP}) with projected augmented wave pseudopotentials \cite{kresse96_iterative_schemes, kresse96_efficiency_total_energy, kresse_joubert_paw, blochl_paw_94}. The revised Perdew-Burke-Ernzerhof functionals for solids (PBEsol) \cite{PBEsol} were used to relax the material structure. The valence electron configurations for Mn, Ga and N in the potentials used are given by 3$p^6$4$s^2$3$d^5$, 4$s^2$3$d^{10}$4$p^1$ and 2$s^2$2$p^3$, respectively. The plane wave cut-off energy was set at 800 eV for optimal convergence.

In order to simulate the AFM ground state, the magnetic spins at manganese sites were ordered in a non-collinear $\Gamma^\text{5g}$ pattern, as illustrated in Figure \ref{fig:MGN_unitcell}. A $2\times2\times2$ MGN super cell was set up. For relaxations, the K point mesh was set to a $4\times4\times4$ point grid. For density of states (DOS), a denser K mesh of $8\times8\times8$ points was used. For high accuracy when evaluating octahedral rotations, the threshold condition during ionic relaxation for the forces between ions was set to EDIFFG = $10^{-5}$ eV/\AA. The energy difference threshold in the electron structure relaxation was set to dE $= 10^{-8}$ eV. \\


These input parameters produced a relaxed structure with cubic symmetry, characterized by a lattice constant of 3.771 \AA. \textcolor{black}{The experimentally measured value at 0\degree C is 3.886 \AA\, \cite{mgn_consts}.} Previous computational studies reported values of 3.86 \AA\, \cite{MGN_lukashev}, 3.864 \AA\, \cite{anomalous_hall_effect_mgn} or 3.867 \AA\, \cite{mgn_sto_sciadv}, using PBE \cite{PBE} functional. The local magnetic moments (LMM) per Mn site converged to 1.91 $\mu_B$. Studies based on the PBE functional predicted a value of about 2.40 $\mu_B$ \cite{MGN_lukashev, anomalous_hall_effect_mgn}. The computed LMMs depend on the projected sphere radius and thus lower LMMs are consistent with the smaller lattice constants achieved using PBEsol. \textcolor{black}{The Hubbard U correction produces unwanted localization effects and overestimation of magnetic moments in this compound and is therefore not employed.} \\

Isotropic biaxial strain was applied by stretching or compressing the supercell, \textcolor{black}{constraining the in-plane lattice constants $a$ and $b$, but not the ion movements within the cell.} Strain was applied in percent increments from -5\% to +5\%. In order to evaluate possible octahedral rotations due to strain, initial octahedral rotations of one degree were introduced into the relaxed cubic structure before applying strain. Specifically, out-of-plane rotation for compressive strain and in-plane for tensile strain. For completeness, the full range of strain levels was applied to both in-plane and out-of-plane, initially rotated structures, \textcolor{black}{in-phase and out-of-phase}. These initial rotations are required to force the structure out of a possible local energy minimum where the N-Mn-N angles are all precisely 180\degree, with the highest symmetry applicable to the space group.

A number of oxide perovskites and metals were simula-\\ted for comparison of electrostatic interactions with those of MGN. \textcolor{black}{Additionally, several antiperovskites were included in this comparison: Mn$_{3}$GaC, Mn$_{3}$CuN and Mn$_{3}$SnN, for their similar composition to MGN and diverse properties, such as giant magnetoresistance, large magnetocaloric effect, near-zero coefficient of thermal resisitivity and anomalous Nernst effect, among others \cite{Mn3GaC_GMR_Kamishima_PRB, Mn3GaC_Ni_GMR_APL_2009, ANE_Mn3SnN, chi_nearly_2001, asano_magnetostriction_2008, fruchart_magnetic_1977}. Ni$_{3}$CuN and Ni$_{3}$MgC exhibit superconductivity at low temperatures \cite{MgCNi3_superconduct_nature_2001, CuNNi3_supercond_2013}. Mg$_{3}$SbN is a semiconductor interesting for photovoltaic applications \cite{heinselman_thin_2019, chi_new_2002}.} Oxide perovskites La\textsubscript{0.7}Sr\textsubscript{0.3}MnO\textsubscript{3} (LSMO), SrVO\textsubscript{3} and SrRuO\textsubscript{3} were chosen for their metallic ground states \cite{LSMOstrain1999, onoda_metallic_1991, sro_rondinelli_spaldin_2008_prb}. Transition metals iron, copper, silver and nickel were selected as references for their metallic bonding with and without ferromagnetic order. Si and Ge were selected for their covalent bonding, and NaCl and MgO for their ionic bonding. A few other perovskite\\ oxides were simulated due to their similar structure, but different bonding chemistry from intermetallic antiperovskites: Cubic SrTiO\textsubscript{3}, belonging to the same space group as MGN \cite{sto_strained_ferroel} and orthorhombic manganates CaMnO\textsubscript{3} and SrMnO\textsubscript{3}, as well as orthorhombic scandate DyScO\textsubscript{3} \cite{daoud-aladine_structural_2007, camno3_aschauer_2013_prb, clark_high-pressure_1978}. Simulated structures from Materials Project (MP) \cite{materialsproject} were obtained as starting points. The pseudopotentials used for each atomic type were as per MP recommendations. However, the PBE functionals in\\ the MP models were replaced with their PBEsol equivalents for close consistency with the MGN simulations. Spin-polarized computations were performed on compounds containing magnetic elements. For metals, pseudopotentials that treat $p$ electrons as valence states were used. The MP models were modified to achieve the same thresholds and numerical accuracy as for MGN. Orthorhombic CaMnO\textsubscript{3} was simulated based on Ref. \cite{camno3_aschauer_2013_prb}. Simulated LSMO data were used from Ref. \cite{Hallsteinsen_moreau_LSMO_LFO_PRB_2016}.

Phonon dispersions were computed using the finite displacement method, as implemented in \textit{phonopy} \cite{phonopy_method}. Processing of DOS and band structure calculations was performed using \textit{VASPKIT} \cite{vaspkit}. Henkelman group's \textit{VTST tools} \cite{vtsttools} were used to compute Bader charges. All crystal structure visualizations are produced using \textit{VESTA} \cite{VESTA}.

\section{\label{sec:effect_of_strain_on_struct}Effect of strain on magnetic and structural phases}

\begin{figure}[ht]
    \centering
    \includegraphics[trim={0 0.8cm 0 0.5cm},clip, width=\linewidth]{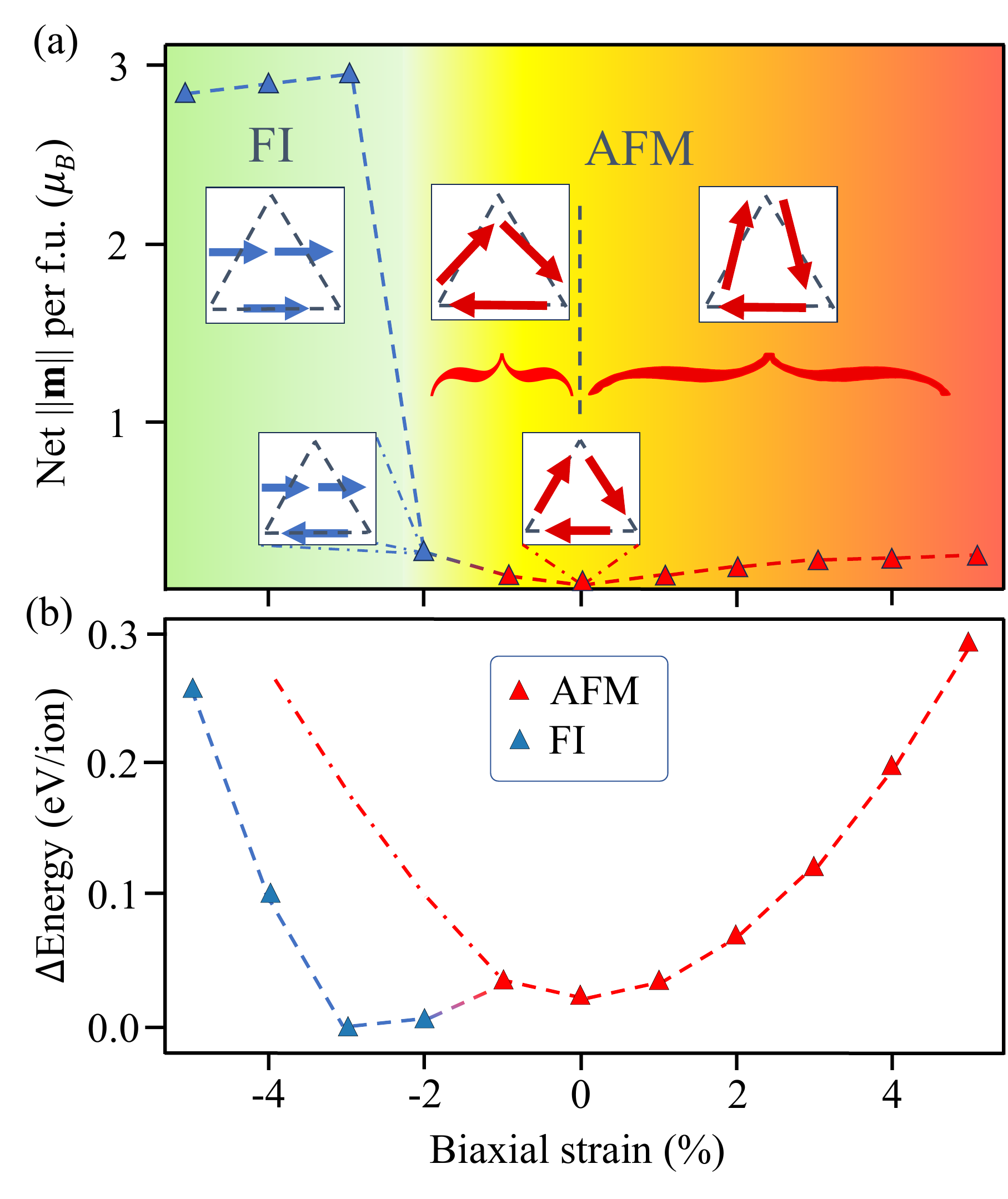}
    \caption{(a) Magnetic phase diagram of MGN showing strain domains that belong to the following phases: (green) parallel ferrimagnetic (FI), (yellow) compressive and (orange) tensile strain AFM phases. Insets show the parallel and antiparallel spin textures found in the FI phase, and the canted and equilateral $\Gamma^\text{5g}$ textures found in the AFM phase. The relative vector lengths in insets are not to scale. Net magnetic moment at different strain levels specified in units of $\mu_B$ per f.u. \textcolor{black}{(b) Relative total energies of strained models converging into AFM (red) and FI (blue) phases.}}
    \label{fig:mag_phase_diagram}
\end{figure}

Applying biaxial strain produced the magnetic phase diagram shown in Figure \ref{fig:mag_phase_diagram} with the magnetic moments tabulated in Table \ref{tbl:magnetic_moments_strain}. The relaxed structure is characteri-\\zed by spin vectors organized in an equilateral triangle in the (111) plane, as illustrated in the inset at 0\% strain in Figure \ref{fig:mag_phase_diagram}, yielding a zero net magnetic moment.

\begin{table*}[ht]
\caption{Magnetic moments at various strain levels (\%): Net scalar moment per f.u. ($||\mathbf{m}||$) and moment vectors $\mathbf{m}_1$, $\mathbf{m}_2$, $\mathbf{m}_3$ of Mn\textsubscript{1}, Mn\textsubscript{2} and Mn\textsubscript{3} lattice sites, respectively, per f.u. All in units of $\mu_B$. The three last columns are the angles $\alpha$, between $\mathbf{m}_2$ and $\mathbf{m}_3$, $\beta$ and $\gamma$, between $\mathbf{m_1}$ and $\mathbf{m_3}$, and between $\mathbf{m}_1$ and $\mathbf{m}_2$, respectively. $\alpha_\text{c}$ is the analogue to $\alpha$ for ions in the underlying (111) crystallographic plane (see Figure \ref{fig:magmom_comp_tensile}). All angles in degrees (\degree).}
	\centering
        \setlength{\tabcolsep}{0.825em}
	\begin{tabular}{cccccc}
		\toprule 
		Strain (\%) & $||\mathbf{m}||$ $\mu_B$& $(\mathbf{m}_1, \mathbf{m}_2, \mathbf{m}_3)$ & $\alpha$ & $\beta$, $\gamma$ & $\alpha_\text{c}$ \\ 
		\midrule
		-5.0 & 2.771  & (-0.540, -0.540, 0.000), (-0.714, -0.715, 0.001), (-0.705, -0.705, -0.001) & 0.12 & 0.06 & 57.98 \\
		-4.0 & 2.820  & (-0.585, -0.585, 0.000), (-0.707, -0.706, -0.001), (-0.701, -0.702, 0.001) & 0.14 & 0.07 & 58.53 \\
		-3.0 & 2.916  & (-0.642, -0.642, 0.000), (-0.710, -0.710, 0.000), (-0.710, -0.710, 0.000) & 0.00 & 0.00 & 59.05 \\
		-2.0 & 0.199  & (1.413, 1.414, -0.001), (-0.745, -0.745, -0.001), (-0.810, -0.809, -0.001) & 0.04 & 0.08 & 59.12 \\
		-1.0 & 0.085 & (1.329, 1.329, 0.000), (-0.134, -1.255, 1.121), (-1.255, -0.134, -1.121) & 71.16 & 54.42 & 59.54 \\
		\, 0.0 & 0.000  & (1.349, 1.349, 0.000), (0.000, -1.349, 1.350), (-1.349, 0.000, -1.350) & 59.98 & 60.01 & 60.00\\
		\, 1.0 & 0.052  & (1.403, 1.403, -0.001), (0.054, -1.420, 1.475), (-1.420, 0.054, -1.475) & 56.28 & 61.86 & 60.42 \\
		\, 2.0 & 0.100 & (1.455, 1.455, 0.000), (0.101, -1.485, 1.586), (-1.485, 0.101, -1.586) & 53.48 & 63.26 & 60.83 \\
		\, 3.0 & 0.125 & (1.500, 1.501, 0.001), (0.132, -1.546, 1.681), (-1.547, 0.136, -1.679) & 51.78 & 64.11 & 61.24 \\
		\, 4.0 & 0.192 & (1.472, 1.472, 0.000), (0.215, -1.552, 1.766), (-1.551, 0.216, -1.767) & 47.17 & 66.42 & 61.86 \\
		\, 5.0 & 0.175 & (1.567, 1.567, 0.000), (0.205, -1.648, 1.854), (-1.648, 0.205, -1.854) & 48.40 & 65.80 & 62.05 \\
		\bottomrule
	\end{tabular}
	\label{tbl:magnetic_moments_strain}
\end{table*}

\begin{figure}
    \centering
    \includegraphics[trim={0 0.8cm 0 0.5cm},clip,width=\linewidth]{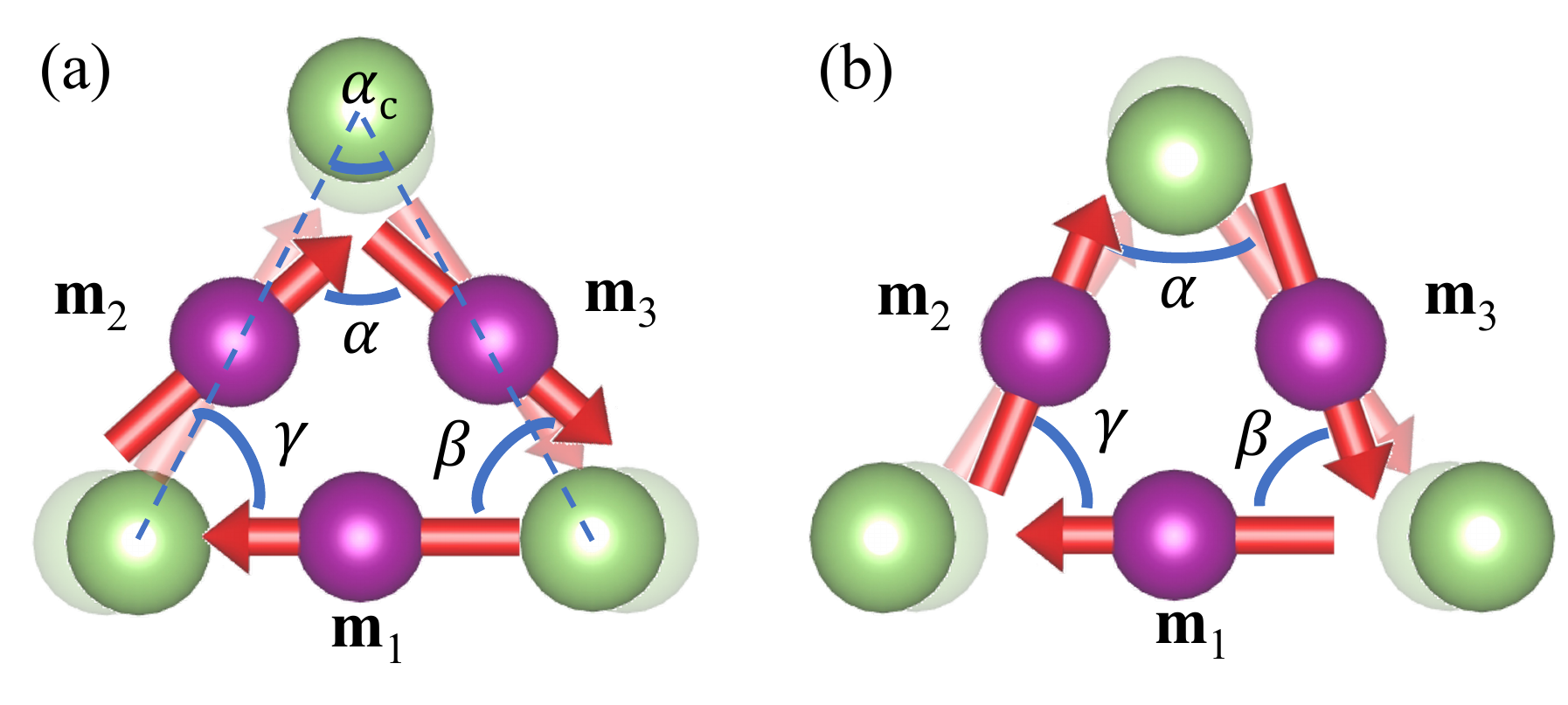}
    \caption{In-plane (111) spin canting under (a) compressive and (b) tensile biaxial (001) strain, as measured by angles $\alpha$, $\beta$ and $\gamma$. $\alpha_\text{c}$ is analogous to $\alpha$, but for the angle between Ga ions in the underlying (111) crystal plane. Canting angles, vector magnitudes and Ga ion displacements are not to scale. Color coding of atomic types as introduced in Figure \ref{fig:MGN_unitcell}. N ion omitted for clarity.}
    \label{fig:magmom_comp_tensile}
\end{figure}

Under biaxial compressive strain, the triangular spin structure is initially preserved. However, a non-zero net magnetic moment emerges in the direction of applied strain, [$\bar{1}\bar{1}0$]. From Table \ref{tbl:magnetic_moments_strain}, it is seen that the biaxial compression of the $ab$ plane results in a distortion of the crystalline (111) plane. This is indicated by a decrease in angle $\alpha_\text{c}$ between the imaginary lines connecting the Ga ions in Figure \ref{fig:magmom_comp_tensile}a. The Ga ion lattice along the (111) plane is no longer made up of equilateral triangles due to the ionic Ga displacements that distort the cubic phase, causing a structural transition into a tetragonal phase under strain. The spin structure is distorted in an opposite way, increasing the angle $\alpha$ between spin vectors $\mathbf{m}_2$ and $\mathbf{m}_3$, and decreasing angles $\beta$ and $\gamma$ in Figure \ref{fig:magmom_comp_tensile}a. The effective spin canting counteracts the lattice distortion, making the net magnetic moment lower as compared to the case if the spin structure followed the crystal distortion. At -1\% strain, this yields a net magnetic moment of 0.1 $\mu_B$ per f.u., with a canting of 11.6\degree\, relative to the crystal lattice. Under greater compressive strain, ferrimagnetic (FI) phases are observed, \textcolor{black}{accompanied by an energy discontinuity in Figure \ref{fig:mag_phase_diagram}b, where they deviate from the (red) AFM phase curve}.
At -2\%, the canting results in an anti-parallel ferrimagnetic order, as shown in the inset in Figure \ref{fig:mag_phase_diagram}a, signified by a sharp increase in net magnetic moment. From Table \ref{tbl:magnetic_moments_strain} we see that in the FI domain, the out-of-plane spin vector component falls sharply to zero, constraining the spin vectors to the $ab$ plane. At -3\% the FI phase has all spins parallel along [$\bar{1}\bar{1}0$], in agreement with the strain tensor. For compressive strain beyond -3\% a small canting of spin vectors occurs, decreasing the net magnetic moment per f.u. from the maximum observed value at -3\%.\\

Biaxial tensile strain in the $ab$ plane produces a non-zero net magnetic moment pointing along [110]. The displacement of Ga ions along the crystalline (111) plane, as shown in Figure \ref{fig:magmom_comp_tensile}b, causes the angle $\alpha_\text{c}$ to increase. The appearance of a net magnetic moment in response to tensile epitaxial strain is consistent with previous studies \cite{mn3ga_volume_strain, MGN_lukashev, MGN_magnetism_caloric_effects, carbides_nitrides_band_structure} that show positive net magnetic moments for MGN under hydrostatic pressure. Figure \ref{fig:magmom_comp_tensile}b illustrates the response of the spin vectors to tensile strain in the (111) plane. From Table \ref{tbl:magnetic_moments_strain}, biaxial stretching of the $ab$ plane results in a distortion of the crystalline (111) plane, opposite to that seen during compressive strain, with angle $\alpha_\text{c}$ now increasing. Analogously to compressive strain, the spin structure responds by canting the spins opposite to the lattice distortion, decreasing angle $\alpha$ and increasing $\beta$, $\gamma$. The spin vectors increase in magnitude with tensile strain, but canting counteracts the net magnetic moment increase.

\begin{figure}[ht]
   \centering
   \includegraphics[width=\linewidth]{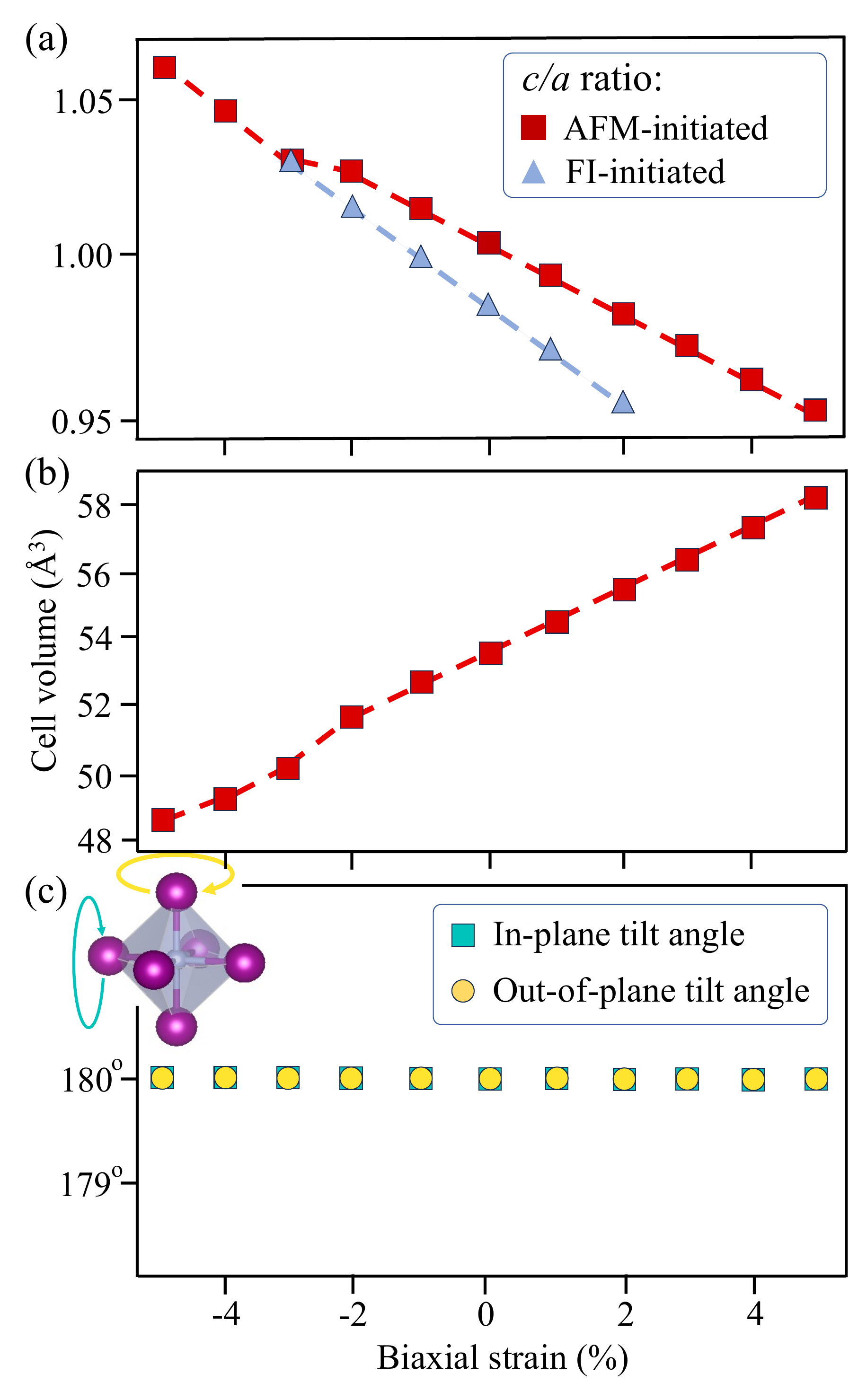}
   \caption{(a) $c/a$ ratio versus strain. Red points are derived from applying strain to the AFM phase ground state. Blue -- from initiating the AFM-phase models with the stable FI-phase (at -3\%) spin structure. (b) Unit cell volume versus strain. (c) N-Mn-N angles as a measure of octahedral tilt at different strain levels.}
    \label{fig:ca_volume_angles}
\end{figure}

Figure \ref{fig:ca_volume_angles}a presents the $c/a$ lattice constant ratios at the strain levels considered. At all levels of biaxial strain considered, a tetragonal unit cell with space group $P4/mmm$ was obtained. The cell volume in Figure \ref{fig:ca_volume_angles}b is monotoni-\\cally increasing throughout the whole range of strain le-\\vels. Both plots show a discontinuity at -2\%, the strain level resulting in the FI-AFM phase transition discussed previously, with the $c$ lattice constant sharply decreasing\\ between -2\% and -3\%. At strain levels above -2\%, this corresponds to a Poisson's ratio of 0.14, while higher compressive strain below -2\% yields a ratio of 0.25. By imposing a parallel FI-phase without spin canting also at strain levels between -2\% and +2\%, initializing the mo-\\dels using the equilibrium spin vectors at -3\%, the $c/a$ ratio follows the Poisson's ratio of 0.25 indicated by blue triangles in Figure \ref{fig:ca_volume_angles}a and extends the trend seen between -3\% and -5\%. This indicates a clear coupling between the spin structure and the resulting unit cell volume.

\begin{table}[ht]
    \caption{Mn-N interatomic distances in MGN as a function of strain in the $ab$ plane, along the $a$ and $c$ axes. Negative strain values indicate compressive strain and positive values signify tensile strain. Zero-strain values indicate the interatomic distances in the relaxed structure.}
    \centering
    \begin{tabular}{p{3cm} p{1.5cm} p{1.5cm}}
	\toprule
	& \multicolumn{2}{c}{Mn-N distance along} \\
	Biaxial strain (\%) & $a$ (\AA) & $c$ (\AA) \\
	\midrule
	-5.0 & 1.791 & 1.903 \\
	-4.0 & 1.810 & 1.892 \\
	-3.0 & 1.828 & 1.881 \\
	-2.0 & 1.847 & 1.897 \\
	-1.0 & 1.866 & 1.892 \\
        \,\,0.0 & 1.886 & 1.886 \\ 
	\,\,1.0 & 1.904	& 1.880 \\
	\,\,2.0 & 1.923	& 1.874 \\
	\,\,3.0 & 1.942	& 1.870 \\
	\,\,4.0 & 1.960	& 1.864 \\
	\,\,5.0 & 1.980	& 1.859 \\
	\bottomrule
    \end{tabular}
    \label{tbl:interatomic_distances}
\end{table}

Next, the effect of biaxial strain on the Mn octahedral rotations is investigated.
Figure \ref{fig:ca_volume_angles}c presents the in-plane and out-of-plane N-Mn-N angles, as a measure of octahedral tilt and rotation in MGN. Under both compressive and tensile strain, using in-plane and out-of-plane initial rotations, the N-Mn-N angles are all within $180\pm0.001$ degrees. That is, the Mn\textsubscript{6}N octahedra do not tilt or rotate about any axis and the initially imposed rotations return back to $a^0a^0a^0$, which is consistent with space group $P4/mmm$. Thus, the magnetic phase transition between $\Gamma^\text{5g}$ and FI is not due to a change in octahedral rotations. 
The forces from strain only result in an elongation and shortening of the bonds. As a proxy for bond lengths, the Mn-N interatomic distances along the $a$ and $c$ axes that resulted from strain in the $ab$ plane are summarized in Table \ref{tbl:interatomic_distances}. The interatomic distances along the $c$ axis are generally decreased as the material is stretched in the $ab$ plane. The only exception is the discontinuity between -2\%  and -3\%. The interatomic distance along the $c$ axis here is equal to half of the lattice constant $c$. This is consistent with the discontinuity seen in the $c/a$ ratio, Figure \ref{fig:ca_volume_angles}a.

\begin{figure}[ht]
    \centering
    \includegraphics[width=0.96\linewidth]{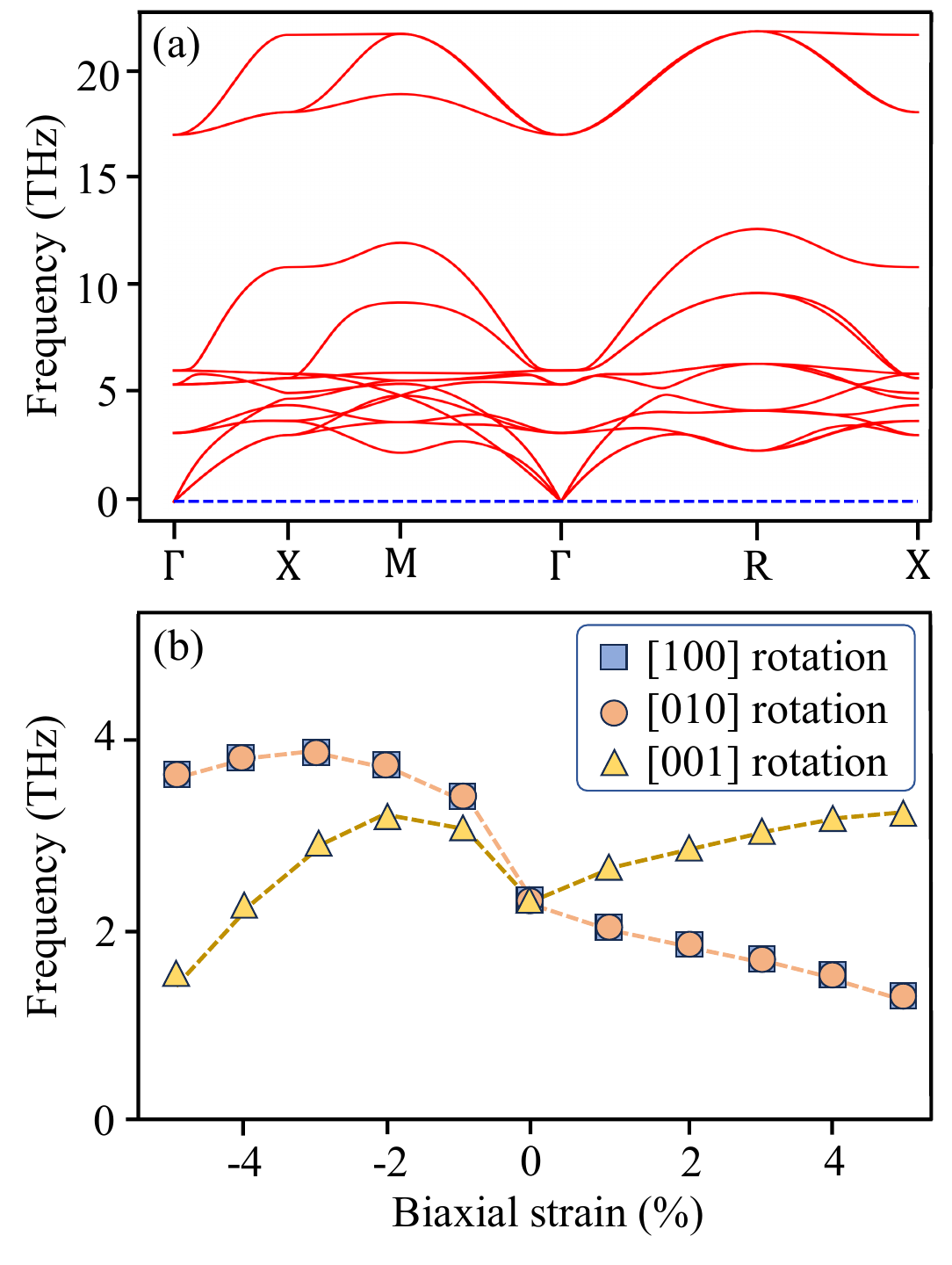}
    \caption{(a) Phonon dispersions at the MGN ground state. (b) R\textsubscript{25} rotational phonon modes in MGN versus strain.}
    \label{fig:phonons_plots}
\end{figure}

In order to study the structural stability of MGN under strain, phonon dispersions were computed along the K path through high-symmetry points \textGamma, X, M, \textGamma, R, X. Figure \ref{fig:phonons_plots}a depicts the phonon dispersion for the cubic ground state. \textcolor{black}{To investigate possible structural distortions occuring by way of octahedral tilts, the evolution of R$_{25}$ acoustic phonon modes under strain is shown in Figure \ref{fig:phonons_plots}b. These three modes correspond to the component degrees of rotational freedom of the octahedra about the three [001] axes.}

\textcolor{black}{The tetragonal phases appearing under the considered levels of strain were found to be dynamically stable.} No unstable R\textsubscript{25} rotational phonon modes are observed in MGN as the phonon frequencies at the R point are always finite. This agrees well with the calculated N-Mn-N angles in Figure \ref{fig:ca_volume_angles}c, showing no octahedral rotations. The change with strain of the $R_{25}$ phonon modes does follow a similar pattern to oxide perovskites, with the [001] rotational mode softening during compression, while the [100] and [010] rotational modes soften during tensile strain \cite{lao_001_111_strain_moreau, strain_phonons_moreau_2017, marthinsen_2018_goldstonelike_phonons, adv_mat_spaldin_rondinelli_2011}.

\section{Effect of strain on the electronic structure}

\begin{figure}[ht]
    \centering
    \includegraphics[width=\linewidth]{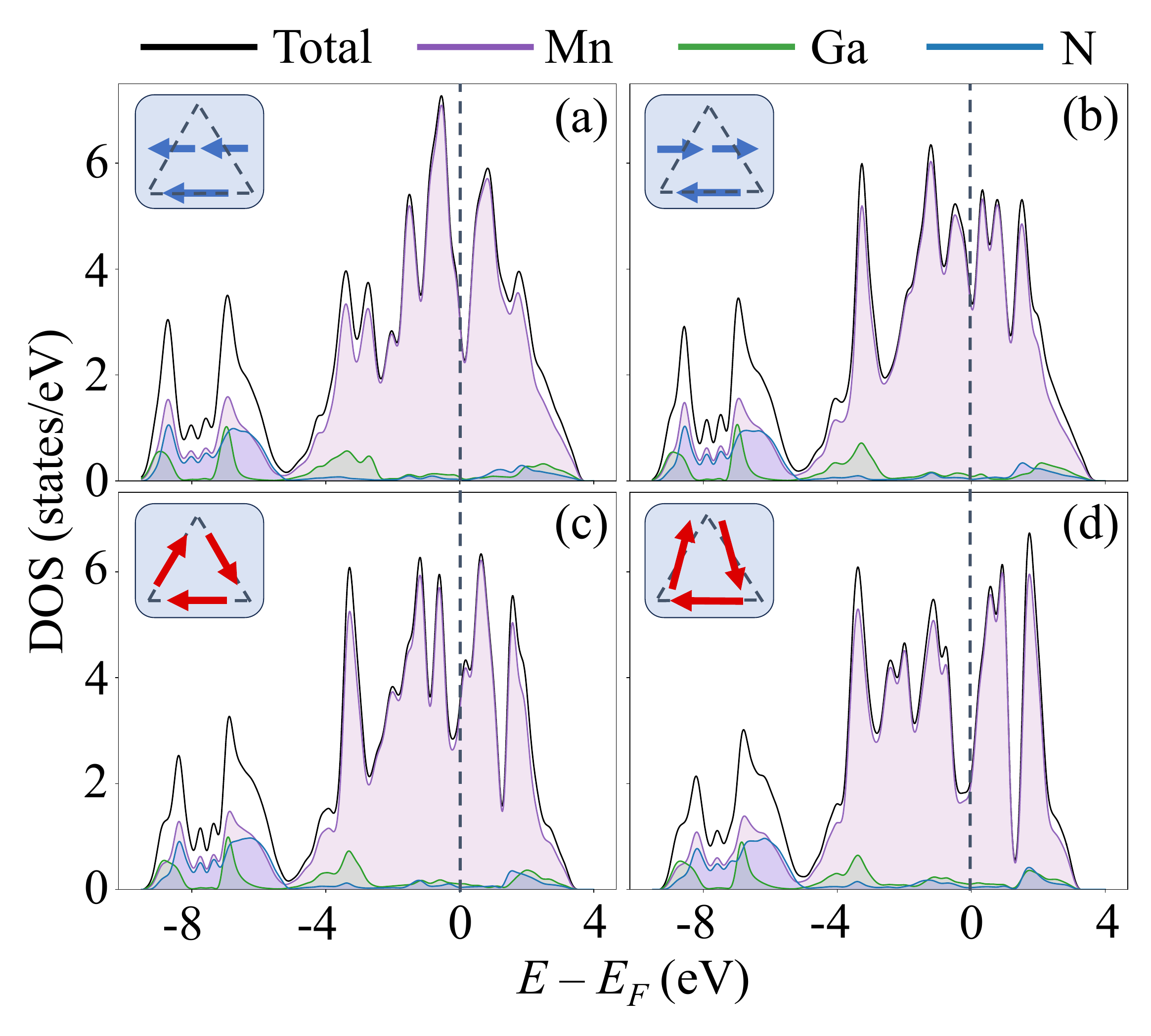}
    \caption{Total MGN DOS and partial DOS for each atomic type at (a) -3\%, (b) -2\%, (c) 0\% (relaxed structure) and (d) 3\% strain, illustrating the electronic structure in the resulting magnetic phases (insets).}
    \label{fig:DOS}
\end{figure}

Figure \ref{fig:DOS} presents the density of states of MGN under the four magnetic phases discussed previously: a) parallel-spin FI at -3\%, b) anti-parallel FI at -2\%, c) AFM \textGamma\textsuperscript{5g} for cubic MGN and d) canted AFM \textGamma\textsuperscript{5g} under 3\% tensile strain. At the Fermi level, the total DOS is dominated by the Mn $d$ states. Ga and N contribute mainly to binding states below the Fermi level. At -2\% strain, Figure \ref{fig:DOS}b indicates that the electronic structure of the transitional anti-parallel FI phase is closer to the AFM phases in Figure \ref{fig:DOS}c and Figure \ref{fig:DOS}d, as compared to the parallel FI phase plotted in Figure \ref{fig:DOS}a. A noticeable change in overall state density distribution, as characterized by shifts in the peaks in the interval (-4, 4) eV, takes place during the AFM-FI transition. However, the visible changes are predominantly observed in the partial density of Mn states and do not provide enough information about charge density transfer between the different ions.

To gauge the transfer of charge between the ions, Bader charges of the constituent ions in MGN are tabulated in Table \ref{tbl:bader}. At 0\% strain, each of the three Mn lattice sites donates an equal amount of charge to N as the two others. With strain, an asymmetry between the contributions of the apex Mn1 site and the equatorial Mn2 and Mn3 sites emerges. However, the net charge transfer between the sites does not change significantly at any strain level, including the AFM-FI transition. On average, the N anion attracts about 1.5 e of charge from the Mn cations under all strain levels considered.

\begin{table}[ht]
	\caption{Bader charges of ions in MGN and in the oxide perovskite STO}
	\centering
	\begin{tabular}{cccccccccc}
		\toprule
		& \multicolumn{8}{c}{Bader charges (e)} \\
            & \multicolumn{5}{c}{MGN} & & \multicolumn{3}{c}{STO} \\
		Strain, \% & Ga & Mn1 & Mn2 & Mn3 & N & \,\, & Sr & Ti & O \\
		\midrule
            -5 & \, 0.05 & 0.41 & 0.48 & 0.48 & -1.42 & & 1.52 & 1.85 & -1.16 \\
            -4 & \, 0.04 & 0.43 & 0.48 & 0.48 & -1.43 & & 1.52 & 1.85 & -1.15 \\
            -3 & \, 0.03 & 0.45 & 0.48 & 0.49 & -1.44 & & 1.53 & 1.84 & -1.14 \\
		-2 &  -0.02 & 0.47 & 0.51 & 0.51 & -1.46 & & 1.53 & 1.86 & -1.12 \\
            -1 &  -0.07 & 0.51 & 0.52 & 0.52 & -1.48 & & 1.53 & 1.85 & -1.11 \\
            \,\,0 &  -0.08 & 0.53 & 0.53 & 0.53 & -1.51 & & 1.60 & 2.07 & -1.22 \\
		\,\,1 &  -0.08 & 0.53 & 0.53 & 0.53 & -1.51 & & 1.55 & 1.84 & -1.08 \\
            \,\,2 &  -0.13 & 0.53 & 0.54 & 0.54 & -1.49 & & 1.55 & 1.82 & -1.07 \\
            \,\,3 &  -0.15 & 0.53 & 0.56 & 0.56 & -1.50 & & 1.56 & 1.79 & -1.06 \\
		\,\,4 &  -0.18 & 0.53 & 0.57 & 0.57 & -1.51 & & 1.56 & 1.81 & -1.04 \\
		\,\,5 &  -0.15 & 0.56 & 0.56 & 0.56 & -1.53 & & 1.56 & 1.81 & -1.03 \\
		\bottomrule
	\end{tabular}
	\label{tbl:bader}
\end{table}

\begin{figure*}[ht]
    \centering
    \includegraphics[width=7in]{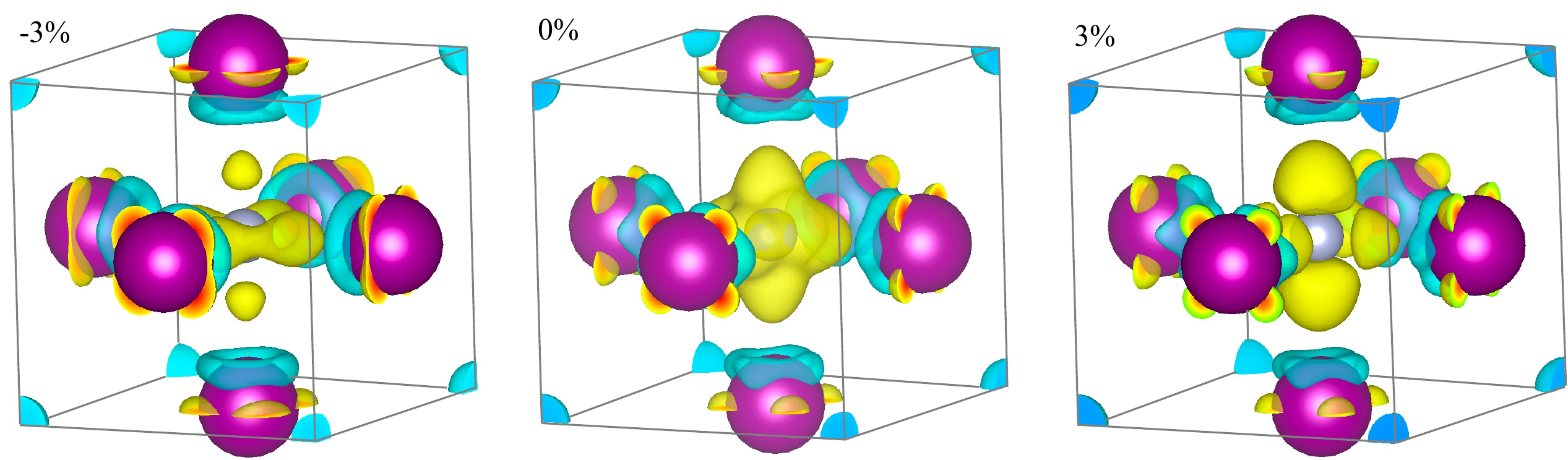}
    \caption{Charge density difference plots of a MGN unit cell at select strain levels: -3\%, 0\% (relaxed) and 3\%. An increase in charge density is marked yellow, while a decrease -- blue. The Ga ions at the corners of the unitcell are hidden so that their contribution can be seen more clearly.}
    \label{fig:MGN_chg_density}
\end{figure*}

Figure \ref{fig:MGN_chg_density} illustrates the charge density transfers qualitatively at -3\%, 0\% and 3\% strain, respectively. It shows Mn sites mainly as charge donors (blue charge density) and N mainly as the acceptor (yellow charge density). During biaxial compression, the equatorial Mn sites donate more charge to N than the apex site Mn1. Under tensile strain the difference in contribution between the Mn sites is less clear, being equal at some strain levels, as seen from Table \ref{tbl:bader}.

\begin{figure}[H]
    \centering
    \includegraphics[width=0.9\linewidth]{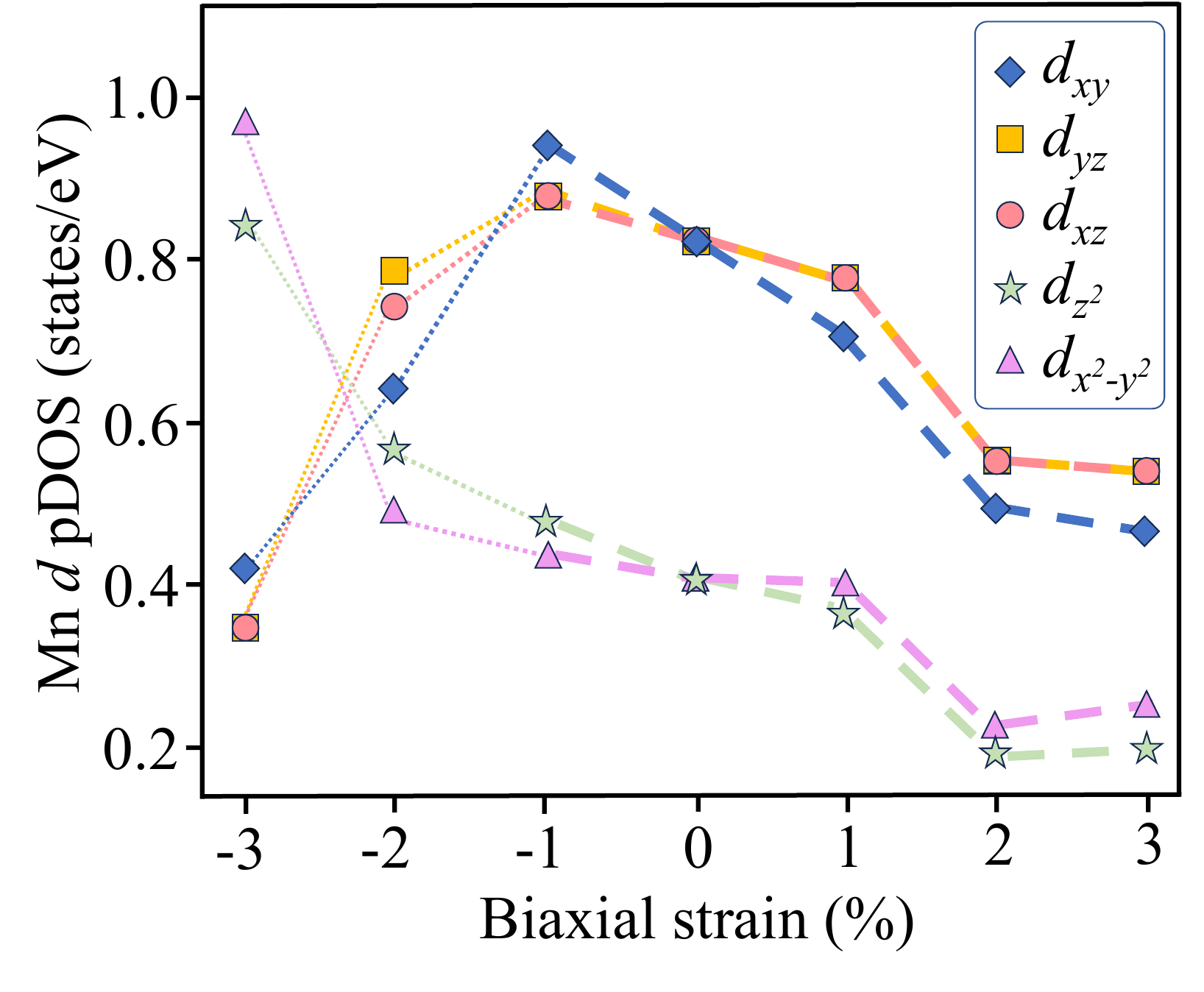}
    \caption{Evolution of the Mn $d$ suborbital partial densities of state at Fermi level with strain.}
    \label{fig:Mn_d_pDOS}
\end{figure}

The discussion up to this point has established that the interionic charge transfer does not change significantly with strain. It is then instrumental to look in more detail at the changes within the Mn $d$ states that dominate the DOS in Figure \ref{fig:DOS}. Figure \ref{fig:Mn_d_pDOS} shows the change in Mn $d$ orbital contribution at the Fermi level with strain. Under 0\% strain the $t_{2g}$ states are threefold degenerate and $e_g$ states are twofold degenerate. Tensile strain decreases all the Mn $d$ suborbital state densities at the Fermi level consistently with how the total DOS evolves in Figure \ref{fig:DOS}. As the crystalline axes in both the cubic and tetragonal phases of MGN coincide with Cartesian $x,y,z$ coordinates, the suborbitals are here labelled using the latter for a more intuitive description. The tetragonal distortion compresses the structure in the $z$ direction,  breaking the orbital degeneracies. This behavior is similar to that observed for the A and B site cation 3$d$ states in oxide perovskites when strained in the (001) plane \cite{d_states_LTO_LVO, spinels_d_states_2005, lao_001_111_strain_moreau}. For MGN, tensile strain does not alter the ordering of the suborbitals with respect to one another. In the case of compressive strain a similar, but opposite, breaking of orbital degeneracies is seen. Starting from the relaxed structure, an increase in the partial and total DOS at Fermi level is observed. At the AFM-FI transition, however, the $d_{xy}$, $d_{yz}$ and $d_{xz}$ state densities are decreasing while the suborbital densities of $d_{x^2-y^2}$ and $d_{z^2}$ states increase. This results in an inversion of the densities of $e_g$ versus $t_{2g}$ states in the FI phase.

To further explore the difference in response to strain from oxide perovskites, Table \ref{tbl:bader} also presents simulated Bader charges of the perovskite oxide SrTiO\textsubscript{3} (STO) per lattice site with strain. This compound is structured according to the same cubic space group as MGN and with a similar lattice constant, having seen use as a substrate for MGN thin film growth \cite{mgn_sto_2013, mgn_sto_sciadv}. STO exhibits $a^0a^0c^-$ rotations in its tetragonal phase \cite{glazer1972} when strained in the (001) plane, making for an interesting comparison with MGN regarding their internal bonding and their chemi-\\stry. The value at 0\% strain represents the computed relaxed cubic structure of STO, while the values at all non-zero strain levels are from the computed tetragonal $I4/mcm$ phase. The reported oxygen Bader charge is here averaged over the three O sites and increases under larger compressive strain. This is consistent with the equatorial O sites shifting closer to the Ti cation and hence accepting more charge from the Ti ion at the B site. However, the total charge transfer summed over the three O sites is significantly higher, 3.0 - 3.6 e, under all strain levels, as compared to the total cations-anion charge transfer in MGN. In fact, computations performed on several other perovskite oxides with other structures -- mostly manganates, for the purpose of comparing with compounds where Mn is at the B site -- revealed Bader charges of 1.1-1.3 e per O site. Similarly to STO, this results in a total charge transfer of about 3.3-3.6 e from the A and B site cations to the O anions, which is more than twice the charge transfer occurring in MGN.

\begin{figure}[ht]
    \centering
    \includegraphics[width=1.06\linewidth]{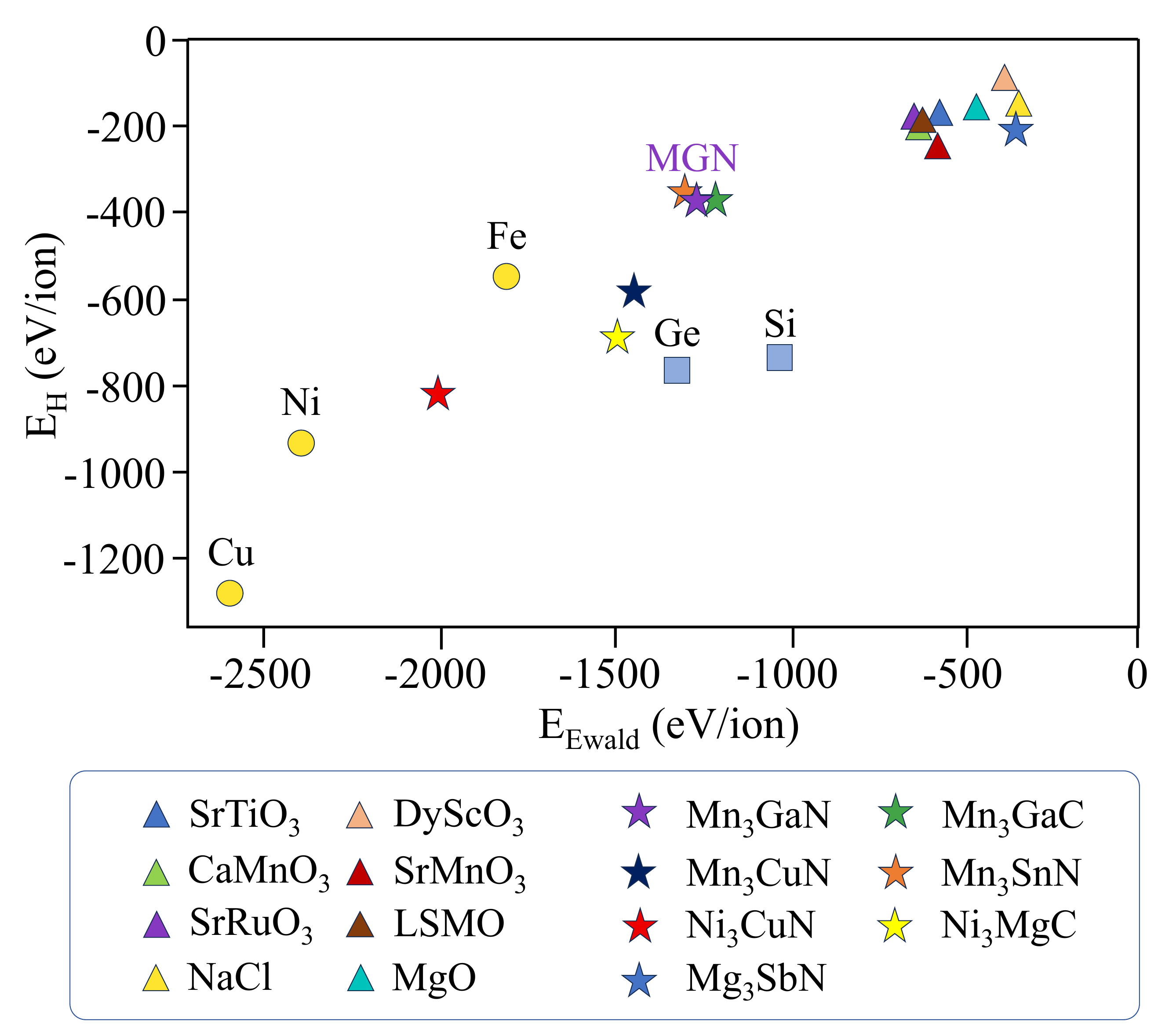}
    \caption{Electrostatic energies of select ionic compounds (triangles), semiconductors (squares), metals (circles), \textcolor{black}{and antiperovskites (stars)} plotted alongside those of MGN.}
    \label{fig:Ewald_hartree}
\end{figure}

Figure \ref{fig:Ewald_hartree} depicts the electrostatic energies of MGN \textcolor{black}{alongside those of a few other antiperovskites and oxide perovskites with diverse properties. Furthermore, the plot includes several ionic salts, covalent semiconductors and metals as references for their respective types of bonding.} The Ewald energy on the $x$-axis is taken as a measure of the ion-ion interactions, while the Hartree energy on the $y$-axis represents the electron-electron interactions. \textcolor{black}{Unlike the oxide perovskites, whether metallic or insulating, antiperovskites do not form a cluster with similar electrostatics. Instead, Figure 9 shows a large variation in bonding within the class of antiperovskites.}

The ionic salts and oxide perovskites, including metallic LSMO and SrRuO\textsubscript{3}, have larger electrostatic energies than MGN, while metals are lower on the energy scale. This reinforces the notion that metal oxide perovskites are more ionic in their bonding than MGN, as the discussion on Bader charges already showed.

In the more ionic oxide perovskites, the O site atoms do not show suborbital reordering due to filled $p$ states and ionic bonding with the B site cation. For the less ionic MGN, strain is accommodated by a redistribution of charge between the Mn $d$ suborbitals, rather than between Mn $d$ orbitals and the N $p$ orbitals. Thus, epitaxial strain does not induce octahedral rotations.

\section{\label{sec:conclusion}Conclusion}
The antiperovskite MGN shows a different response to biaxial strain from that observed in oxide perovskites. When applying biaxial epitaxial strain, both tensile and compressive, the mechanism of accommodating strain-induced phase transitions by way of octahedral rotation or tilt is not seen in MGN. Rather, the octahedra stay in the Glazer tilt system $a^0a^0a^0$ even at high levels ($\pm 5\%$) of biquadratic strain. The electronic structure suggests that this is mediated by a higher degree of bond covalency than seen in perovskite oxides. This allows the bonds to be stretched or compressed to accommodate the strain forces, altering the distribution of electronic states within the cation bands, rather than the interband redistribution mediated by the anion.

Furthermore, the study investigated a magnetic phase transition between AFM and FI spin structures by way of compressive strain on the system, where the state densities of $d_{xy}$, $d_{yz}$ and $d_{xz}$ suborbitals, and those of $d_{x^2 - y^2}$ and $d_{z^2}$, become reversed at the Fermi level. The FI phase produces a discontinuous decrease in cell volume, but preserves the same space group symmetry as in the AFM phase.

\section{Acknowledgments}
The Norwegian Metacenter for Computational Science is acknowledged for providing computational resources at Uninett Sigma 2, project no. NN9301K.

\bibliography{apssamp}

\end{document}